# Surface stress effects on the electrostatic pull-in instability of nanomechanical systems

**Hamed Sadeghian**

Department of Optomechatronics, Netherlands Organization for Applied Scientific Research, TNO, Stieltjesweg 1, 2628 CK, Delft, The Netherlands
Tel.: +31 (0)88 866 43 55,
E-mail: hamed.sadeghianmarnani@tno.nl

**Abstract:** The electrostatic pull-in instability (EPI), within the framework of the nanoelectromechanical systems (NEMS) has been shown as a robust and versatile method for characterizing mechanical properties of nanocantilevers. This paper aims to investigate the surface effects, specifically residual surface stress and surface elasticity, on the EPI of micro and nano-scale cantilevers as well as double clamped beams. Since the cantilever has one end free, it has no residual stress, thus the strain-independent component of the surface stress or intrinsic surface stress has no influence on the EPI, as long as it has small deformation. The strain-dependent component of the surface stress or surface elasticity changes the bending stiffness of the cantilever and, consequently, induces shifts in the EPI. For double clamped beams, the effective residual surface stress comes into play and modifies the effective residual stress of the beam. The nonlinear electromechanical coupled equations, which take into account the surface effects are solved numerically. The theoretical results presented in this paper indicate that the EPI is very sensitive to the surface effects, especially when a double clamped beam is employed. The results show that the influence of surface effects on the EPI of cantilevers become more profound when the thickness is below 50 nm, while the influence on double clamped beams is significant even at sub-micron scale. The present study can provide helpful insights for the design and characterization of NEMS switches. Moreover, the results can be used to provide the proof of concepts of a new surface stress sensing method using EPI in nanomechanical sensor systems.

## 1. Introduction

Surface stress has a great impact on a wide range of surface-related phenomena, such as surface reconstruction [1], phase transformation [2, 3, 4], epitaxial growth [5], and self-assembled domain patterns [6, 7]. Due to the very high surface-to-volume ratio in micro and nanomechanical structures such as cantilevers, double clamped beams, nanowires and nanotubes, the role of surface stress effects on the mechanical properties are very significant and thereby have been extensively studied by several researchers using experimental measurements [8, 6, 9, 10, 11, 12, 13] and theoretical investigations (through both atomistic simulations [14, 2, 15, 16, 17] and modifications to continuum theory [18, 19, 20, 16] as explanations for the observed size effects at ultra-small scales. One of the most commonly used experimental approach to measure the surface stress effects is bending of a micro/nanocantilever arising from the changes in the surface stress, which was first proposed by Stoney [8]. Later, this method has been shown as the most appropriate method for work in liquid environment [21, 22, 23]. The other technique is to monitor changes in the resonance frequency of cantilevers or double clamped beams, and has been shown as the best suited for use in gaseous or vacuum environment [24]. This was first reported by Lagowski et al. [25], who studied the resonance frequency of GaAs cantilevers as a function of surface preparation and ambient atmosphere. They proposed a one-dimensional model in which the surface stress was replaced by a compressive axial force. Later on, Gurtin et al. [26] showed that this model is incorrect and when is corrected, yields a resonance frequency that is independent of surface stress. In order to explain the experimental results, carried out by Lagowski, the effects of surface elasticity or strain-dependent surface stress on the stiffness, and, consequently on the resonance frequency of cantilevers have been examined by many researchers [26, 27, 28, 29, 20, 30]. However, they concluded that the effect on the resonance frequency of micro/sub-micron cantilevers is negligible. As an example, Lu et al. [27] theoretically showed that on a clean silicon cantilever of 1 μm thickness, one finds shift in the resonance frequency of about 10 ppm. in practical purposes, the mass loading effects are the same order of magnitude or larger than the surface stress effects and, therefore, it is almost impossible to distinguish between shift in the resonance frequency due to surface effects and that of the loading mass, thus sensing the changes in surface parameters with the use of shifts in the resonance frequency is extremely difficult, if not impossible. Therefore, the lack of a sensitive, accurate and reliable sensing method for measuring the surface stress parameters exists. Since the surface effects influence the stiffness of nanomechanical structures, a sensing method, which is solely based on the changes in the stiffness can be very useful in studying the surface effects.

Recently, Sadeghian *et al.* [30] experimentally demonstrated the use of electrostatic pull-in instability (EPI), within the framework of the nanoelectromechanical systems (NEMS), for the characterization of micro/nano suspended structures such as cantilevers. If a voltage is applied between the cantilever and a fixed electrode, which is separated from the cantilever by a dielectric medium, the beam deflects toward the fixed electrode. Once the voltage exceeds a critical value, an increase in the electrostatic force becomes greater than the mechanical restoring force, resulting in an instable behavior (collapsing of the beam to the fixed electrode) known as the electrostatic pull-in instability phenomenon. The uniqueness of the pull-in method lies in its well-known sharp instability and the possibility of applying a force distributed along the length of the beam. The measurement is independent of mass-loading effects and the method-induced error is the lowest among all characterizing methods in NEMS [30]. More recently, the application of EPI on sensing adsorbate stiffness in nanomechanical resonators has been presented [31, 32]. Because the method is very sensitive to the changes in the elastic behavior, while being independent of mass loading effects, it can be used for studying the surface effects.

The EPI phenomenon is also very important in Microelectromechanical systems based on the electrostatic actuation such as microwave variable capacitors [33], MEMS switches [34, 35], Digital micromirrors [36, 37] and microvalves for fluidics applications [38], where the size of the system is in the range of hundreds of micrometer. At this scale the surface effects can be reasonably neglected. However, recent experimental demonstrations reported the development of NEMS switches with thicknesses and gaps smaller than 50 nm [39, 40]. They offer very low pull-in voltages and also simultaneously offer very short switching times in the sub-microsecond range. Like MEMS switches, in NEMS switches determination of the EPI is critical in the design process. Since the characteristic sizes of these devices shrink to nanometers, thus exhibiting inherently large surface to volume ratio, surface effects may play a crucial role in their behavior and specifically on their EPI phenomena.

Up until now, extensive studies have been carried out on the investigation of the EPI, but the investigation of the surface effects on the EPI is rare. Using the generalized Young-Laplace equation and implementing it in Euler-Bernoulli beam model, Bryan Ma *et al.* [41] investigated the surface effects on a cantilever that is electrostatically actuated. Fu *et al.* studied the size effects on the electrostatic pull-in instability of nanobeams at the presence of surface energies. In this paper, the surface effects on the EPI of both cantilevers and double clamped beams are investigated. The effects of various geometrical and physical parameters such as thickness, length and residual stress at the presence of the surface effects are investigated. Due to the nonlinearity of the elastic-electrostatic interaction, exact analytical solutions are generally not available. For this, the generalized differential quadrature (GDQ) algorithm is employed to solve the nonlinear differential equation [42]. The results demonstrates the application of EPI on the study of surface effects. In the first part of the paper, the electromechanical coupled models of the cantilever and the double clamped beam subjected to an applied voltage are presented, in which, the effects of surface stress and surface elasticity are taken into account. The second part of the paper, is the results and discussion.

## 2. Surface elastic properties: An overview

Atoms at or near the surfaces experience different environment and lack some of the atomic neighbors present in the bulk state. Consequently, the energy of these atoms are different from that of the atoms in the bulk. This excess energy associated with the surface atoms is called surface energy [43]. The lagrangian description of the surface energy is given by [44]

$$\gamma = \frac{1}{A_0} \sum_{n=1}^{\infty} (U^{(n)} - U^{(0)}) \qquad (1)$$

where $U^{(n)}$ is the total energy of the atom n under the area $A_0$ and $U^{(0)}$ is the total energy of an atom in a perfect lattice far away from the free surface. In contrast to the surface energy, surface stress is defined as the forces which oppose an elastic deformation of the surface and changes the interatomic distance at a constant number of atoms [45, 46]. The surface stress tensor can be determined from the change in the surface energy due the deformation of the area, thus, the change in the surface energy is equal to the work done by the surface stress in deformation of the area through an infinitesimal elastic strain $d\varepsilon_{\alpha\beta}$

$$d(\gamma A_0) = A_0 \sigma^s_{\alpha\beta} d\varepsilon_{\alpha\beta} \qquad (2)$$

where $\sigma^s_{\alpha\beta}$ is the surface stress tensor and is defined as

$$\sigma^s_{\alpha\beta} = \frac{d\gamma}{d\varepsilon_{\alpha\beta}} = (\tau_0)_{\alpha\beta} + S_{\alpha\beta\lambda\kappa} \qquad (3)$$

where $\tau_0$ is the residual surface stress or intrinsic surface stress, and sometimes is called "strain-independent" p of the surface stress [26, 27]. The surface elasticity tensor, S, can be defined as

$$S_{\alpha\beta\lambda\kappa} = \frac{d^2\gamma}{d\varepsilon_{\alpha\beta}d\varepsilon_{\lambda\kappa}} \quad (4)$$

The surface elastic properties can be determined from atomistic calculations [47, 48].

## 3. Nonlinear distributed electromechanical coupled model incorporating surface effects

When a driving voltage is applied between the electrodes (cantilever or double clamped beam as movable electrodes and the substrate as a fixed electrode), the electrostatic pressure deflects the movable electrode. The mechanical bending strain energy $U_b$ of the bulk is given by

$$U_b = \int_0^L \int_A \frac{1}{2}\sigma_{xx}\varepsilon_{xx}dAdx \quad (5)$$

where $A$ is the cross-sectional area of the uniform beam, $L$ is the length of the beam, $\sigma_{xx}$ is the axial stress, and $\varepsilon_{xx}$ is the normal strain, respectively. Using the linear strain-displacement relation

$$\varepsilon_{xx} = \frac{\partial u}{\partial x} = -z\frac{d^2y}{dx^2} \quad (6)$$

we obtain

$$U_b = \int_0^L \int_A \frac{E}{2}\left(-z\frac{d^2w}{dx^2}\right)dAdx = \int_0^L \frac{EI}{2}\left(\frac{d^2w}{dx^2}\right)^2 dx \quad (7)$$

where u and w are in-plane and trasverse displacement components of the midplane, $E$ is the Young's modulus, and $I$ is the effective moment of inertia of the cross section, respectively. The transverse coordinate is z and the midplane coincides with $z = 0$. The total surface energy, $U_s$, at the entire beam surface is

$$U_s = 2\int_s (\gamma(\varepsilon) - \gamma(0))ds = 2\int_s \frac{1}{2}S\varepsilon^2 ds = \int_s Sbt^2\left(\frac{d^2w}{dx^2}\right)^2 dx \quad (8)$$

where $S$ is the surface elasticity, b is the width of the cantilever, s is the surface area of the beam and t is the thickness of the cantilever, respectively. Residual stress, due to the inconsistency of both the thermal expansion coefficient and the crystal lattice period between the substrate and thin film, is unavoidable and the residual force can be expressed as [49]

$$T_r = \sigma_r(1-\nu)bt \quad (9)$$

where $\sigma_r$ is the biaxial residual stress (equal to zero for the cantilever beams), and $\nu$ is the Poisson's ratio. In double clamped beams, the intrinsic residual stress, or the strain-independent surface stress $\tau_0$ modifies the residual stress, therefore, the energy stored in the beam due to the residual stress and strain-independent surface stress is.

$$\begin{aligned} U_r &= \int_0^L T_r\left(\frac{dw}{dx}\right)^2 dx \\ &+ \int_0^L 2b\tau_o\left(\frac{dw}{dx}\right)^2 dx \\ &= \int_0^L (1-\nu)b\left(t\sigma_r \right. \\ &\left. + \frac{2\tau_0}{(1-\nu)}\right)\left(\frac{dw}{dx}\right)^2 dx. \end{aligned} \quad (10)$$

When a beam is in tension, the actual beam length $\acute{L}$ is longer than the original length L. In double clamped beams, although there is no displacement in the x-direction at the beam ends, the bending of the beam generates an axial force, i.e.,

$$T_a = \frac{EA}{L}(\acute{L} - L) \approx \frac{Ebt}{2L}\int_0^L \left(\frac{dw}{dx}\right)^2 dx \quad (11)$$

The surface elasticity, or strain-dependent surface stress modifies the axial force in the double clamped beam, consequently, the effective axial force is written as

$$\tilde{T}_a \approx \left(\frac{Ebt}{2L} + \frac{2Sb}{L}\right)\int_0^L \left(\frac{dw}{dx}\right)^2 dx \quad (12)$$

and the energy stored in the beam due to axial force is expressed as

$$\begin{aligned} U_a &\approx \int_0^L \left(\frac{Ebt}{2L}\left(4\frac{S}{Et}\right. \right. \\ &\left.\left. + 1\right)\int_0^L \left(\frac{dw}{dx}\right)^2 dx\right)\left(\frac{dw}{dx}\right)^2 dx \end{aligned} \quad (13)$$

The electrical co-energy $U_e^*$, which is the sum of the electrostatic energy stored between the upper and lower electrode of the beam and the electrostatic energy of the voltage source is given by:

$$U_e^* = \frac{1}{2}\int_0^L \frac{\epsilon_r\epsilon_o bV^2}{(g-w(x))}dx \quad (14)$$

the total potential energy U of the system is

$$U = U_b + U_s - U_r - U_a + U_e. \quad (15)$$

The variation of total energy is zero at the equilibrium position, i.e.,

$$\delta U = \delta U_b + \delta U_s - \delta U_r - \delta U_a + \delta U_e = 0 \quad (16)$$

Therefore, the nonlinear integro-differential equation of a double clamped beam subjected to an applied voltage with surface effects can be written as

$$EI(24\lambda + 1)\frac{d^4w}{dx^4}$$
$$-\left(bt\sigma_r(1-v)(1+2\beta)\right.$$
$$\left.+\frac{Ebt}{2L}(4\lambda+1)\int_0^L\left(\frac{dw}{dx}\right)^2 dx\right)\frac{d^2w}{dx^2} \quad (17)$$
$$=\frac{\epsilon_r\epsilon_o bV^2}{2(g-w(x))^2}(1+f_f)$$

where non-dimensional parameters $\lambda$ and $\beta$ are defined as

$$\lambda = \frac{S}{Et} \text{ and } \beta = \frac{\tau_0}{(1-v)\sigma_r t} \quad (18)$$

and

$$f_f = 0.65\left(\frac{g-w(x)}{b}\right) \quad (19)$$

is the first-order fringing field correction. Cantilevers have no residual and axial stress, thus, the nonlinear differential equation is written as

$$EI(24\lambda+1)\frac{d^4w}{dx^4}$$
$$= \frac{\epsilon_r\epsilon_o bV^2}{2(g-w(x))^2}(1+f_f) \quad (20)$$

In cantilevers, only surface elasticity changes the bending stiffness, but in double clamped beams, both surface elasticity and surface stress come into play. One can get the traditional electromechanical coupled equation of cantilevers and double clamped beams [42], without surface effects if $\lambda = \beta = 0$.

## 4. Results and Discussion

The numerical examples of a cantilever and a double clamped beam made of single crystal silicon subjected to a voltage are now presented. The generalized differential quadrature method (GDQM) [42] is used to transform the aforementioned nonlinear integro-differential equations into the corresponding discrete forms and the NewtonRaphson method is implemented for solving the set of nonlinear algebraic equations that result from the application of GDQM. The material properties are given as follows [50]: Young's modulus $E$ is 169 GPa, Poissons ratio is 0.06, and the permittivity of air is 8.85 pF/m. The width is kept constant and is 4 μm.

Figure 1 shows $\frac{|\Delta V|}{V}(\%)$ as a function of surface elasticity S and thickness, where ΔV is the shift in the EPI due to the surface effects and V is the EPI without the surface effects. It can be seen that reducing the thickness of cantilever results in a bigger change in the EPI for a constant S. For cantilevers thinner than 50 nm, the EPI is sufficiently sensitive to the surface elasticity and, therefore, in practice, it is possible to track the changes in the surface elasticity due to either molecular adsorption [51, 20] or surface reconstruction [52, 20, 16]. The bending of the cantilever subjected to a voltage is also influenced by the surface elasticity. According to equation 8, negative/positive surface elasticity reduces/increases the bending stiffness of the cantilever, and that changes the bending behavior. This is shown in figure 2. The inset shows the non-dimensional difference of the end tip of the cantilever versus non-dimensional parameter $\lambda$. The cantilever modeled in figure 2 is 10 nm thick and 10 μm long.

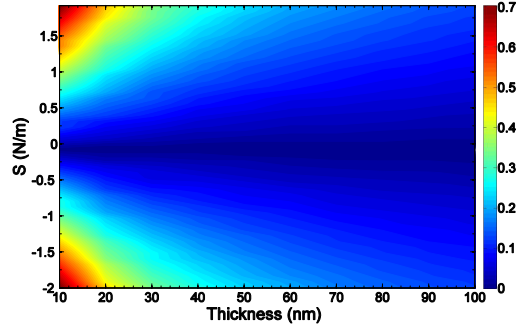

**Fig. 1.** Non-dimensional changes in pull-in voltage, $\frac{|\Delta V|}{V}(\%)$, of cantilever due to surface elasticity for various thicknesses.

Figure 3. (a)-(f) shows the results for double clamped beams. Figure 3. (a) shows the $\frac{|\Delta V|}{V}(\%)$ for various lengths and thicknesses. $\tau_0$ and S are kept as 1 N/m [27]. The initial gap g is 1 μm and the residual stress $\sigma_r$ is assumed zero. One can learn from the figure that increasing the surface to volume ratio (reducing the thickness), while keeping the lateral dimensions constant, results in a higher shifts in the EPI. Increasing the length would cause increase in the EPI as well. Moreover, it can be seen that even for thicker double clamped beams, i.e. 1 μm, the shift in the EPI is significant. This is due to the fact that unlike cantilevers, in double clamped beams the residual surface stress $\tau_0$ modifies the residual stress in the beam, because the beam is clamped in both sides and therefore it cannot release the stress. It has been shown that the residual stress

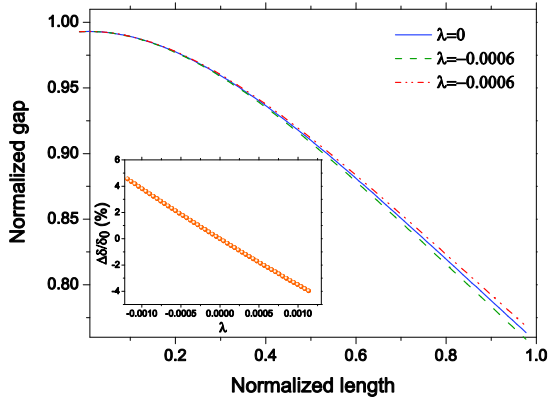

**Fig. 2.** Normalized deflection of a 10 nm thick silicon cantilever for different . The inset shows relative difference between the end tip of the cantilever incorporating surface elasticity and the end tip of the cantilever with no surface elasticity

is the most dominant effects on the EPI [42], and, consequently, the residual surface stress has a great impact on the EPI of double clamped beams.

Figure 3. (b) shows the effects of $\tau_0$, while varying the thickness. $\tau_0$ is varying from negative (compressive surface stress) to positive values (tensile surface stress). As it is expected, reducing the thickness would increase the sensitivity of the EPI to $\tau_0$. Longer double clamped beams also show the same, see figure 3. (c). The results are demonstrated for a 100 nm thick beam and $\sigma_r$ is zero. As discussed above, the residual stress, which is exhibited due to fabrication, plays a substantial role for the EPI [42, 53]. The residual stress, $\sigma_r$ can be either compressive (negative) or tensile (positive). The compressive one causes buckling of the beam, reducing the gap between the beam and the substrate, and consequently, decreasing the pull-in voltage. Negative $\tau_0$ induces the same. This is shown in figure 3. (d). As it is shown in equations 8 and 13, the surface elasticity S modifies the bending stiffness as well as the axial force. The negative S causes softening of the beam and the positive one induces a stiffening effect. Figures 3. (e) and 3. (f) show the effect of S on the pull-in voltage, while changing the thickness and the length of the double clamped beam. increasing the surface to volume ratio (figures 3. (e)) increases the sensitivity of $\frac{|\Delta V|}{V}$(%) to S. For a fix thickness, scaling the lateral dimension, i.e., changing the length does not influence the $\frac{|\Delta V|}{V}$(%), see figure 3. (f). The bending properties of the double clamped beam, due to applied voltage also depend on the surface effects.

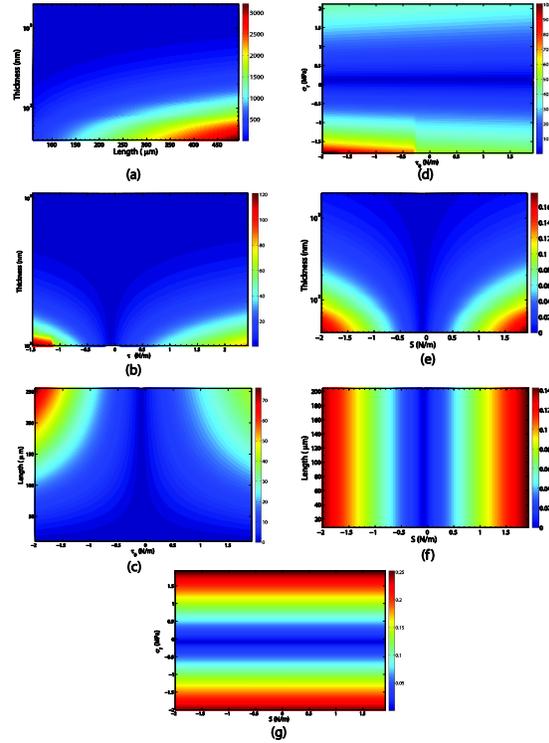

**Fig. 3.** Non-dimensional changes in pull-in voltage, $\frac{|\Delta V|}{V}$(%), of double clamped beam due to surface effects. (a) Effects of the thickness and the length on $\frac{|\Delta V|}{V}$, (b), (c), and (d) Effects of $\tau_0$ in different thicknesses, lengths and residual stress on $\frac{|\Delta V|}{V}$(, (e) and (f) Effects of surface elasticity $S$ on $\frac{|\Delta V|}{V}$ for various thicknesses and lengths. (g) Effects of surface elasticity as well as residual stress.

Under the same voltage, when the surface effects are considered the bending is significantly different from the case of no surface effects. This is shown in figure 4. Unlike cantilevers, not only the surface elasticity S influences the bending, but also the residual surface stress $\tau_0$.

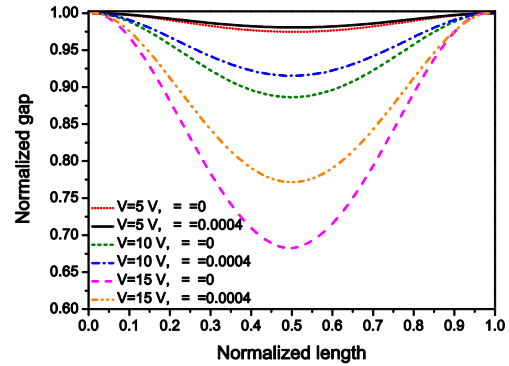

**Fig. 4.** Deflection of a 500 nm thick, 100 μm long double clamped beam subjected to applied voltage. The surface effects induce changes in the bending properties.

## 5. Conclusions

In this paper, the effects of surface stress and surface elasticity on the electrostatic pull-in instability (EPI) of micro/nano cantilevers and double clamped beams were studied. The nonlinear electromechanical coupled model of cantilevers and double clamped beams incorporating the surface effects were developed. The model showed in detail the contribution of residual surface stress and surface elasticity on the bending stiffness, effective residual stress, and the axial stress. A major conclusion is that unlike the resonance frequency method, the EPI is free from mass loading effects, and thus it can be used for quantitative measurements of surface stress and surface elasticity. Compared to resonance frequency measurements, with EPI a relatively thicker structures can be used for surface stress sensing. Moreover, the theoretical results reveal that a high quality factor cantilever or double clamped beam, which is still required and this appears to rule out the use of the resonance frequency method for bio-molecular sensing, is not an issue for the EPI.

## Acknowledgments

This research was financially supported by Early Research Program (ERP) 3D Nanomanufacturing at Netherlands Organization for Applied Scientific Research (TNO).

2006. NEMS '06. 1st IEEE InternationalConference on, 2006, pp. 1117–1120.